# Understanding Individual Differences: Towards Effective Mobile Interface Design and Adaptation for the Blind

Tiago Guerreiro, Hugo Nicolau, João Oliveira, Joaquim Jorge, Daniel Gonçalves
IST/INESC-ID

Av. Professor Cavaco Silva, IST

2780-990, Porto Salvo, Portugal

tjvg@di.fc.ul.pt, hman@vimmi.inesc-id.pt, jmgdo@ist.utl.pt, jaj@acm.org,
daniel.goncalves@inesc-id.pt

## ABSTRACT
No two people are alike. We usually ignore this diversity as we have the capability to adapt and, without noticing, become experts in interfaces that were probably misadjusted to begin with. This adaptation is not always at the user's reach. One neglected group is the blind. Spatial ability, memory, and tactile sensitivity are some characteristics that diverge between users. Regardless, all are presented with the same methods ignoring their capabilities and needs. Interaction with mobile devices is highly visually demanding which widens the gap between blind people. Our research goal is to identify the individual attributes that influence mobile interaction, considering the blind, and match them with mobile interaction modalities in a comprehensive and extensible design space. We aim to provide knowledge both for device design, device prescription and interface adaptation.

## Categories and Subject Descriptors
H.5.2 [Information Interfaces and Presentation]: User Interfaces – *Input devices and strategies, User-centered design.*

## 1. INTRODUCTION
Mobile technology has shown tremendous evolution in the last few years and new devices are constantly presented to consumers. However, the begotten interfaces are still challenging to the disabled user. Particularly, blind users face several difficulties dealing with such increasingly visual-based devices and interfaces. Assistive technologies designed to overcome the limitations imposed by the lack of vision are, to say the least, stereotypical, one-size-fits-all solutions, which neglect the differences amongst the target population. These differences, sometimes irrelevant when in the presence of vision, gain special importance in its absence and should be considered thoroughly.

Our research's ultimate goal is to identify and quantify the individual attributes that make a difference to a blind user when interacting with a mobile device. The mapping between individual capabilities and product demands will then enable us to suggest the best device for a particular individual or inform designers about the most promising methods and attributes, thus promoting inclusive design. Further, only by having a deep understanding of the differences between individuals and how they are related with interface demands, will we be able to provide effective adaptive interfaces. Only by knowing what to account for, we will be able to act accordingly.

## 2. Individual Differences among the Blind
Each mobile phone brand presents us with a set of stylish application-enriched devices created to impress us with its multi-context functionalities. They are not mere communications devices and have become indispensable in our daily lives.

These gadgets are an opportunity for all, however still a challenge for many. Although manufacturers present us with a multitude of devices, interaction methods and characteristics, there is not an understanding of the attributes that maximize each individual user's performance. Indeed, mobile devices are selected accordingly to aesthetic, number and type of applications, and other technical features, disregarding the suitability of the interface to the user.

In particular, disabled users are presented with general assistive technologies that are focused to overcome their main disability. Blind users are now able to operate mobile devices resorting to screen reading software, one that replaces the on-screen visual information by its auditory representation (text-to-speech). However, mobile interfaces are extremely visual and a large amount of information is lost in this visual-audio replacement. Possible examples are the need to resort to tactile capabilities to feel the keypad, cognitive capabilities to memorize letter placement or spatial orientation to have a notion of the device and its components. Visual feedback makes these attributes dispensable or less pertinent. The absence of visual feedback makes them relevant and worthy of consideration.

In this sense, the most relevant individual differences (Figure 1) within the target group were gathered in a previous experiment [1] where we interviewed several professionals (psychologists, occupational therapists, rehabilitation technicians, and teachers) that work daily with blind users. These have shown that individual differences between blind people are likely to have a wider impact on their abilities to interact with mobile devices than among sighted people. Tactile sensitivity, spatial ability, short-term memory, blindness onset age and age are mentioned as deciding characteristics for a blind user mobile performance.

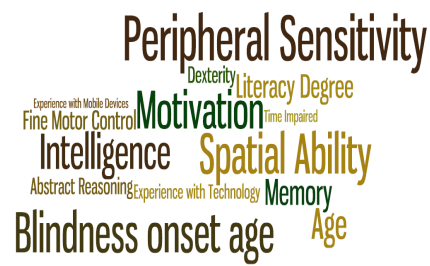

**Figure 1 – Individual attributes relevancy**

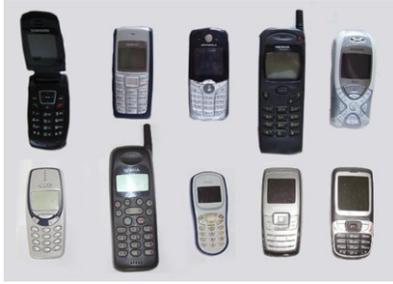

**Figure 2 - Mobile devices used in the evaluation**

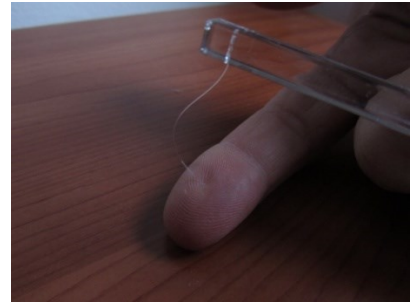

**Figure 3 - Semmes-Weinstein monofilament test**

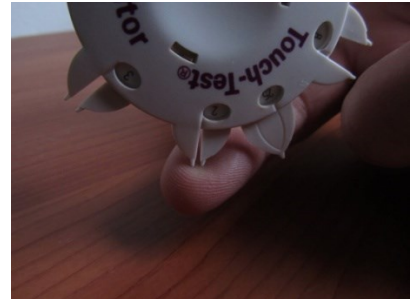

**Figure 4 - Disk-Criminator**

Persad et al. [2] propose an analytical evaluation framework based on the Capability-Demand theory where user capabilities at sensory, cognitive and motor levels, are matched with product demands. We embrace the Capability-Demand theory and strive to understand the relation between user capabilities and the interface demands. This quest can be performed at several levels: hardware, where we can understand the demands imposed by physical components (e.g., spacing, size, relief and contrasting material for the keypad and its keys); task, where we can relate the load imposed by particular methods (e.g., two different text-entry methods); or interaction primitives and parameters (e.g., taps and gestures, and their on-screen locations, on a touch-screen device). The demands are imposed both at the hardware and software levels and both are worth considering.

## 3. Tools for Informed Hardware Design

The first accessibility barrier comes with the hardware design. The physical and immutable interaction challenges are to be addressed before any other as they can be hazardous, despite assistive technologies or adaptations, to the user's effectiveness.

We performed studies with 13 blind people consisting on key acquisition tasks with 10 keypad-based mobile devices (Figure 2). In this study, we intended to understand the impact of individual differences (tactile sensitivity, working memory, attention and spatial ability) among the blind, when interacting with mobile device keypads, as well as how these differences are revealed when confronted with different device demands (key relief, spacing, size, contrasting material and mark). To assess the participants' tactile capabilities, two different components of tactile sensitivity were measured. Pressure sensitivity, was determined using the Semmes-Weinstein monofilament test [3] (Figure 3). In this test, there are several nylon filaments with different levels of resistance, bending when the maximum pressure they support is applied. The other measured component of tactile sensitivity was spatial acuity, using the Disk-Criminator [4] (Figure 4). This instrument measures a person's capability to distinguish one or two points of pressure on the skin surface. The cognitive evaluation focused on two components of cognitive ability: verbal and non-verbal. The verbal component was evaluated in terms of working memory (the Digit Span subtest of the revised Wechsler Adult Intelligence Scale (WAIS-R) was used [5]). The non-verbal component, which consists of abilities independent of language or culture, was evaluated in terms of spatial ability: the human being ability to create and manipulate mental images, as well as maintain orientation relatively to other objects. Spatial ability was measured using the combined grades of the tests Planche a Deux Formes and Planche du Casuiste [6] (Figure 5).

Results showed that different capability levels have significant impact on user performance and that this impact is related with the device and its demands. More than that, the results showed that it is possible to find the best device for each person. Although that is difficult to accomplish, the opposite is not, i.e. finding the most inclusive devices and the acceptable limits for each device characteristic (and its underlying demand).

## 4. Tools for Dynamic Adaptation

The same aforementioned advantages and opportunities present themselves when looking to a software user interface, with a plus: it can be adapted in real time to suit the user's abilities. What we argue here is that to provide fully functional adaptive systems one must understand the variables and dimensions to adapt. Looking to our target group, our studies presented spatial ability, memory and tactile sensitivity as important aspects to consider. Currently, we are undertaking studies with 50+ blind users interacting with several mobile control interfaces (type: keypad, joypad, joystick, touch screen) with their low-level primitives and parameters. With this we expect to be able to provide a model that relates the user's characterization with the device demands. Once again, we are aware that it is not feasible to have an adaptive system relying on clinical trials. However, the knowledge underlying the model encloses an adaptive design space, one that gives researchers and designers the understanding on what are the interface variables and parameters to adapt.

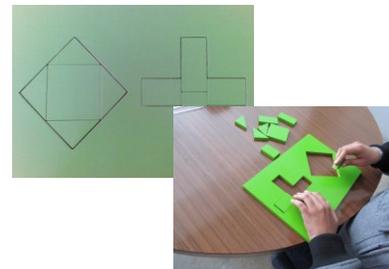

**Figure 5 - Spatial ability assessments**

## 5. Conclusions

Individual differences among the blind have a great impact on the different mobile interaction proficiency levels they attain. General-purpose interfaces and assistive technologies disregard these differences. In this paper, we argue that both the users' capabilities as the device demands should be explored to foster inclusive design. By doing so we will be able to provide more inclusive devices and adapt interfaces accordingly to the variations within the users, maximizing each individual performance.

## 6. ACKNOWLEDGMENTS

We would like to thank the people that have been participating in our studies. This work was partially supported by FCT (INESC-ID multiannual funding) through PIDDAC Program funds. This research was sponsored in part by Portuguese Science Foundation grant MIVIS-PTDC/EIA-EIA/104031/2008. Hugo Nicolau and Tiago Guerreiro were supported by FCT, grants SFRH/BD/46748/2008 and SFRH/BD/28110/2006, respectively.